\documentclass[aps,preprintnumbers,amsmath,amssymb,nofootinbib]{revtex4}  
\usepackage{graphicx} 
\usepackage{bm}
\usepackage{slashed}

\begin{document}
\title{$B\to K^* M_X$ vs $B\to K M_X$ as a probe of a scalar-mediator dark matter scenario}
\author{Alexander Berezhnoy$^{a}$ and Dmitri Melikhov$^{a,b,c}$}
\affiliation{
$^a$D.~V.~Skobeltsyn Institute of Nuclear Physics, M.~V.~Lomonosov Moscow State University, 119991, Moscow, Russia\\
$^b$Joint Institute for Nuclear Research, 141980 Dubna, Russia\\
$^c$Faculty of Physics, University of Vienna, Boltzmanngasse 5, A-1090 Vienna, Austria}
\begin{abstract}
  Recently, Belle II reported the observation of the decay $B\to K M_X$, $M_X$ the missing mass,
  with the branching ratio much exceeding ${\cal B}(B\to K \nu\bar\nu)$ which is the only Standard
  Model (SM) process contributing to this reaction. If confirmed, this might be an indication of new nonSM
  particles produced in this decay.
  One of the possible explanations of the observed effect could be light dark-matter
  (DM) particles produced via a scalar mediator field.
  We give simple arguments, that a combined analysis of the $B\to K M_X$ and
  $B\to K^* M_X$ reactions would be a clean probe of the scalar mediator scenario:
  (i) making use of an observed value ${\cal B}(B\to K M_X)\simeq 5.4\, {\cal B}(B\to K \nu\bar\nu)_{\rm SM}$ and 
  (ii) assuming that the effect is due to the light dark matter coupling to the top quark via
  a {\it scalar} mediator field, one finds an upper limit
  ${\cal B}(B\to K^* M_X) < 2.8 \, {\cal B}(B\to K^* \nu\bar\nu)_{\rm SM}$. 
  Within the discussed scenario, this upper limit does not depend on the mass of the scalar mediator nor on the
  specific details of the unobserved dark-matter particles in the final state.
\end{abstract}
\date{\today}
\maketitle
\normalsize

\noindent {\it 1. Introduction.} A recent Belle II observation \cite{belle} of $B\to K M_X$ (for which 
also the notation $B\to K \slashed{E}$, $\slashed{E}$ the missing energy, is used) 
at the level much exceeding the Standard model (SM) prediction for $B\to K\nu\bar\nu$ \cite{2023damir} 
\begin{eqnarray}
  \label{belle}
{\cal B}(B^+\to K^+ M_X)=(2.3\pm 0.7)\times 10^{-5}\simeq (5.4\pm 1.5)\,{\cal B}(B^+\to K^+\nu\bar\nu)_{\rm SM}, 
\end{eqnarray}
opened a window for immediate discussions of possible new physics effects capable to explain this result
(see e.g. recent publications \cite{2023damir,2023a,2023hiller,2023aa,2023dm,2023b,2023c}).
One of the discussed options is the decay into Dark Matter (DM) particles \cite{2023dm}
with multiple scenarios for the content of these particles and the possible mediators.
The goal of this short communication is to remark that combining the present Belle II
result for the $B\to K M_X$ with the hypothesis of a DM origin of the enhancement
of $\Gamma(B\to K M_X)$ coupled to the SM particles via a
{\it scalar mediator field} leads to interesting and rather robust
constraints independent of further details of the DM model.

\vspace{.3cm}
\noindent 2. {\it The $B\to (K,K^*) \bar\chi\chi$ decays via scalar mediator $\phi$.} 
As an example we consider a simple but rather general Lagrangian describing the iteraction of DM particles $\chi$ with the
top quark via a scalar mediator field $\phi$ \cite{Leffa,Leffb}:
\begin{eqnarray}
 \mathcal{L_{\rm int}} = - \frac{y m_t}{v} \phi\, \bar t t - \kappa \phi \bar \chi \chi \ ,
    \label{eq:lagrangian}
\end{eqnarray}
The corresponding effective Lagrangian describing the flavour-changing neutral current (FCNC) $b\to s\phi$ vertex has the form
\cite{Leffa,Leffb}
\begin{eqnarray}
\label{Leff}
\mathcal{L}_{b \rightarrow s\phi} =  g_{b\to s\phi}\,\phi\, \bar s_L b_R  + {\rm h.c.},  \qquad
  g_{b\to s\phi}=\frac{y m_b}{v} \frac{3 \sqrt{2} G_F m_q^2 V^*_{qs} V_{qb}}{16 \pi^2} 
\end{eqnarray}
This interaction leads to the following amplitudes of the $B\to (K,K^*)\bar\chi\chi$ decay via the $\phi$ mediator: 
\begin{eqnarray}
  A(B(p)\to K(p-q)\bar\chi(k)\chi(q-k))&=&-i\langle K(p-q)\bar\chi(k)\chi(q-k)|L_{b\to s\phi}|B(p)\rangle\nonumber\\
  &=&\langle \bar\chi\chi|\bar\chi\chi|0\rangle \kappa\frac{1}{M_\phi^2-q^2}g_{b\to s\phi}\langle K(p-q)|\bar s_L b_R|B(p)\rangle\\
  A(B(p)\to K^*(p-q)\bar\chi(k)\chi(q-k))&=&-i\langle K^*(p-q)\bar\chi(k)\chi(q-k)|L_{b\to s\phi}|B(p)\rangle\nonumber\\
  &=&\langle \bar\chi\chi|\bar\chi\chi|0\rangle \kappa\frac{1}{M_\phi^2-q^2} g_{b\to s\phi}
\langle K^*(p-q)|\bar s_L b_R|B(p),
\end{eqnarray}
where 
\begin{eqnarray}
\langle \bar\chi\chi|\bar\chi\chi|0\rangle\equiv \langle \bar\chi(k)\chi(q-k)|\bar\chi(0)\chi(0)|0\rangle, 
\end{eqnarray}
and $q$ is the momentum of the outgoing $\bar\chi\chi$ pair of unobserved DM particles, $M_X^2\equiv q^2$ is
the missing mass squared. Using these amplitudes, one can calculate $d\Gamma(B\to (K,K^*)\bar\chi\chi)/dq^2$, where 
the summation over polarizations of the DM particles $\chi$ and the integration over their phase space are
performed leading to 
\begin{eqnarray}
  \sum_{\chi\,{\rm  polar}}|\langle \bar\chi(k_1)\chi(k_2)|\bar\chi(0)\chi(0)|0\rangle|^2
 \delta(m_\chi^2-k_1^2)\delta(m_\chi^2-k_2^2)\theta(k_{10})\theta(k_{20})\delta(q-k_1-k_2) dk_1dk_2=\Pi_{\chi}(q^2).
\end{eqnarray}
We are interested in the ratio of the differential distributions in two
reactions proceedings via the scalar mediator $\phi$: 
\begin{eqnarray}
\label{ratio0}
R^{(\phi)}_{K^*/K}(q^2)=\frac{d\Gamma(B\to K^*\bar\chi\chi)/dq^2}{d\Gamma(B\to K\bar\chi\chi)/dq^2}. 
\end{eqnarray}  
For this ratio, the explicit form of $\Pi_\chi(q^2)$ is of no importance: it is the same function in
$d\Gamma(B\to K\bar\chi\chi)/dq^2$ and $d\Gamma(B\to K^*\bar\chi\chi)/dq^2$ and therefore drops out in the ratio.
It is also obvious that the specific properties of the DM particles $\chi$ are of no importance too:
the final estimate holds for the DM particles independent of their spins. Moreover, the ratio is not sensitive to the
properties of the scalar mediator $\phi$ as its propagator also cancels in the ratio. 
Essential for the ratio (\ref{ratio0}) is merely {\it the operator structure of the vertex}, $\phi\,\bar s_L b_R$. 
Using the QCD equations of motion, it is straigtforward to calculate the amplitudes
\begin{eqnarray}
  \langle K|\bar s_L b_R|B \rangle &=&\frac{1}{2}\langle K|\bar s (1-\gamma_5) b|B\rangle=
  \frac{1}{2}\langle K|\bar s b|B\rangle=
  \frac{1}{2} \frac{M_B^2-M_K^2}{m_b-m_s}f_0^{B\to K}(q^2),\nonumber\\
  \langle K^*|\bar s_L b_R|B \rangle &=& \frac{1}{2}\langle K^*|\bar s(1-\gamma_5) b|B \rangle=
  -\frac{1}{2}\langle K^*|\bar s \gamma_5 b|B \rangle
  =-i (\epsilon q)\frac{M_{K^*}}{m_b+m_s}A_0^{B\to K^*}(q^2),  
\end{eqnarray}
where the dimensionless form factors $f_0$ and $A_0$ are well-known quantities parametrizing the
$\langle K|\bar s \gamma_\mu b|B\rangle$ and $\langle K^*|\bar s \gamma_\mu\gamma_5 b|B\rangle$ amplitudes  
\cite{wsb}. The decay rates of interest then take the form:
\begin{eqnarray}
  \frac{d\Gamma\left(B\rightarrow K \bar\chi\chi\right)}{dq^2} &=& \left|\frac{g_{b\to s\phi}\,\kappa}{M_\phi^2-q^2}\right|^2
  \frac{\lambda^{1/2}(M_B^2,M_K^2,q^2)}{16\pi M_B^3} |\langle K|\bar s_L b_R|B \rangle|^2,\\
 \frac{d\Gamma\left(B\rightarrow K^*\bar\chi\chi\right)}{dq^2} &=& \left|\frac{g_{b\to s\phi}\,\kappa}{M_\phi^2-q^2}\right|^2
  \frac{\lambda^{1/2}(M_B^2,M_K^2,q^2)}{16\pi M_B^3} \sum_{K^*\mbox{polar}}|\langle K^*|\bar s_L b_R|B \rangle|^2. 
\end{eqnarray}
Performing summation over the $K^*$ polarizations, we find for the ratio (\ref{ratio0}): 
\begin{eqnarray}
\label{ratio}
R^{(\phi)}_{K^*/K}(q^2)=
\frac{\lambda^{3/2}(M_B^2,M_{K^*}^2,q^2)}{\lambda^{1/2}(M_B^2,M_K^2,q^2)(M_B^2-M_K^2)^2}
  \left[\frac{m_b-m_s}{m_b+m_s}\right]^2\left|\frac{A_0(q^2)  }{ f_0(q^2)}\right|^2. 
\end{eqnarray}

\begin{figure}[t!]
\begin{center}
\includegraphics[height=4.5cm]{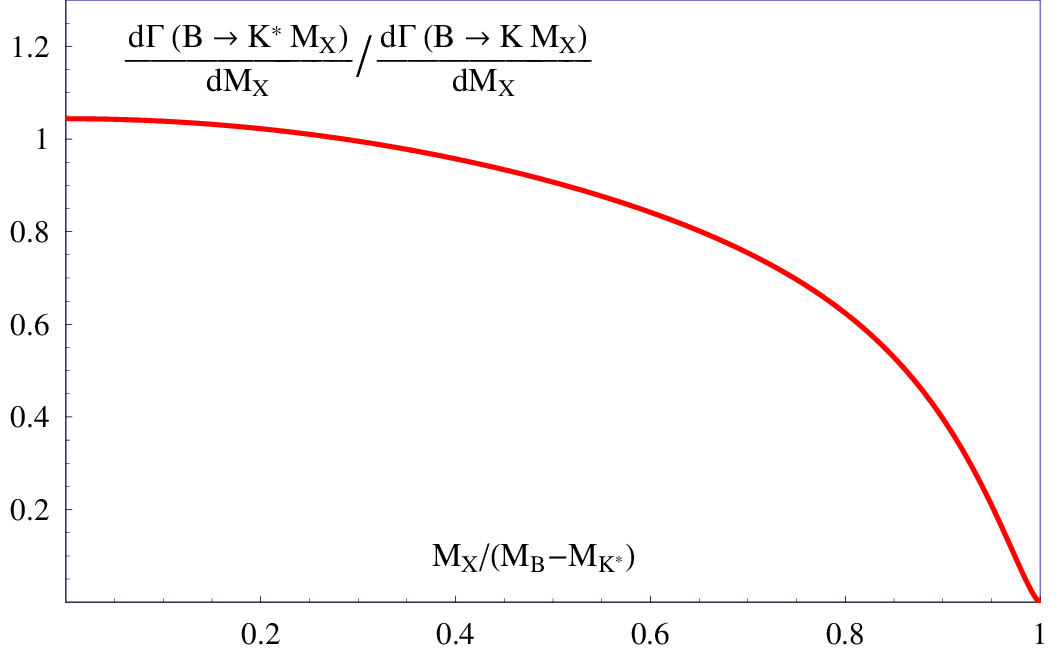}
\caption{\label{Plot:1}The ratio $R^{(\phi)}_{K^*/K}=d\Gamma(B\to K^* M_X)/d\Gamma(B\to K M_X)$ vs $M_X$, $M_X^2=q^2$, $q$ is the 
 momentum of the outgoing $\chi\bar\chi$ pair of DM particles interacting with the top quark via scalar mediator field. 
  Phase-space restriction  
  related to the finite masses of the DM $\chi$ particles should be taken into account
  (i.e. not all values of
  $M_X$ shown in the plot are accessible in the decay). We point out that all factors related
  to the $\phi$-propagator (including $M_\phi$ and possible finite-width effects) and to the details
  of the $\phi\chi\chi$ vertex cancel in the ratio.}
\end {center}
\end{figure}

\vspace{.3cm}
\noindent {\it 3. Numerical estimates.} 
For the form factors, we make use of the results from \cite{ballP} and \cite{ballV}
in the form of convenient parametrizations of \cite{ms2000}: 
\begin{eqnarray}
\label{f0}
f_0(q^2) &=& \frac{0.33}{1-0.7\, r_V+0.27\, r_V^2},\quad r_V=q^2/m_{B_s^*}^2, \\
\label{A0}
A_0(q^2) &=& \frac{0.37}{(1-0.46\, r_P)(1-r_P)},\quad r_P=q^2/m_{B_s}^2. 
\end{eqnarray}
We then come to the following numerical prediction for the ratio of the differential distributions in the reactions
$B\to K \phi\to K\chi\chi$ and $B\to K^* \phi\to K^*\chi\chi$ shown in Fig.~\ref{Plot:1}.
Notice that using more recent parametrizations
for the form factors, e.g., those from \cite{damir2023epjc,hpqcd2023}, leads to the uncertainty of around 2\% in
$R^{(\phi)}_{K^*/K}$. 

\noindent {\it 4. Discussion}. As clear from the plot, one can obtain an upper limit:
\begin{eqnarray}
  \label{upper}
\delta\Gamma(B\to K^* M_X)_\phi< 1.05\, \delta\Gamma(B\to K M_X)_\phi. 
\end{eqnarray}
This upper limit is obtained from Fig.~\ref{Plot:1} at $M_X^2=q^2\to 0$.
The upper limit of Eq.~(\ref{upper}) is a rather conservative
estimate and in practice one may expect a larger suppression of $\delta\Gamma(B\to K^* M_X)_\phi$
compared to $\delta\Gamma(B\to K M_X)_\phi$: The $q^2$-region providing the dominant
contribution to the $B\to (K,K^*)\bar\chi\chi$ cross sections is mainly determined by the resonant structure
of the $\phi$-propagator and corresponds to nonzero $q^2$ thus leading to a further suppression
$\delta\Gamma(B\to K^* M_X)_\phi < \delta\Gamma(B\to K M_X)_\phi$.

Now, assuming that the observed enhancement of $\Gamma(B\to K \nu\bar\nu)_{\rm SM}$ is due to unobserved DM particles 
interacting via a scalar mediator we can write
\begin{eqnarray}
\Gamma(B\to K M_X)&=&\Gamma(B\to K\nu\bar\nu)_{\rm SM}+\delta \Gamma(B\to K M_X)_{\phi}, \\
\Gamma(B\to K^* M_X)&=&\Gamma(B\to K^*\nu\bar\nu)_{\rm SM}+\delta \Gamma(B\to K^* M_X)_{\phi}. 
\end{eqnarray}
For ${\cal B}(B\to K^*\nu\bar\nu)/{\cal B}(B\to K\nu\bar\nu)$ one obtains $2.5\pm 0.7$ making use
of old estimates \cite{mns1997a,mns1997b,mns1998} or $2.23\pm 0.40$ using the recent analysis
of \cite{damir2023epjc}. Within the quoted uncertainties
these two values are in excellent agreement with each other. 

Collecting together the central values of the discussed constraints: 
\begin{eqnarray}
\Gamma(B\to K^*\nu\bar\nu)_{\rm SM}&\simeq&2.5\, \Gamma(B\to K\nu\bar\nu)_{\rm SM}\hspace{0.7cm}\mbox{[Theoretical estimates]}\\
\delta \Gamma(B\to K M_X)_{\phi}&\simeq& 4.4\, \Gamma(B\to K\nu\bar\nu)_{\rm SM}\hspace{0.7cm} \mbox{[Belle measurement]}\\
\delta \Gamma(B\to K^* M_X)_{\phi}&<& 1.05\, \delta \Gamma(B\to K M_X)_{\phi}\hspace{0.4cm} \mbox{[Our result here]}, 
\end{eqnarray}
we then obtain the following upper bound for the $\Gamma(B\to K^* M_X)$
within the {\it scalar mediator scenario}:  
\begin{eqnarray}
\Gamma(B\to K^* M_X) < 2.8\, \Gamma(B\to K^*\nu\bar\nu)_{\rm SM}. 
\end{eqnarray}
Therefore, for the case of scalar mediator, an expected enhancement of
the branching ratio $\Gamma(B\to K^*M_X)$ compared to the SM value
is approximately two times smaller than the enhancement for $\Gamma(B\to K M_X)$, Eq.~(\ref{belle}).  
This constraint is well compatible with the present experimental limit \cite{belle2017}
\begin{eqnarray}
{\cal B}(B\to K^* M_X)\le 2.7\times 10^{-5}\le 2.7\,{\cal B}(B\to K^*\nu\bar\nu)_{\rm SM},
\end{eqnarray}
where at the second step we have used a recently updated estimate ${\cal B}(B\to K^*\nu\bar\nu)_{\rm SM}=
(1.03\pm 0.15)\times 10^{-5}$ \cite{damir2023epjc}. 

We therefore conclude that the combined analysis of ${\cal B}(B\to K^*M_X)$ and ${\cal B}(B\to K\, M_X)$
may provide a useful probe of the scalar mediator scenario [e.g., may exclude this scenario if the obtained upper limit is violated]. 
Since the obtained constraints are not sensitive to the scalar mediator mass $M_\phi$, the latter should then be determined from other
observables.

Before closing this letter, let us briefly comment on the case of the spin-1 ($V$) mediator.
Here the situation is much more involved:
The effective $b\to s V$ Hamiltonian coincides with the $C_{7\gamma}$ part of the $b\to s\gamma$ effective Hamiltonian
of the Standard Model \cite{mns1998} and one readily obtains formulas for the ratio $R^{(V)}_{K^*/K}$ via the $B\to (K,K^*)$
tensor form factors making use of the analytic results given in \cite{mns1998}. Unlike the scalar mediator case,
the effect of the propagator of the vector mediator does not drop out
from the ratio and $R^{(V)}_{K^*/K}(q^2)$ is strongly sensitive to the vector mediator mass. 
For instance, for $M_V\sim 5\;{\rm GeV}$ and at $q^2 \le  5\;{\rm GeV}^2$, one finds $R^{(V)}_{K^*/K}(q^2) > 15$,
and this value rises steeply if $M_V$ decreases. So, the behaviour of $R^{(V)}_{K^*/K}$ is 
qualitatively different from that of $R^{(\phi)}_{K^*/K}$. A detailed analysis of the vector-mediator scenario, however,
lies beyond our interests in the present letter and we leave it for our future analysis.

\vspace{.2cm}\noindent 
{\it Acknowledgments.}
The research was carried out within the framework of the program ``Particle Physics and Cosmology'' 
of the National Center for Physics and Mathematics.



\begin{thebibliography}{100}
\bibitem{belle}
I.~Adachi \textit{et al.} [Belle-II Collaboration], 
  {\it Evidence for $B^+\to K^+\nu\bar\nu$ Decays}, arXiv:2311.14647 [hep-ex].

\bibitem{2023damir}
L.~Allwicher, D.~Becirevic, G.~Piazza, S.~Rosauro-Alcaraz, O.~Sumensari,
{\it Understanding the first measurement of $Br(B\to K\nu\bar\nu)$}, 
arXiv:2309.02246.

\bibitem{2023a}
P.~Athron, R.~Martinez, and C.~Sierra, 
{\it B meson anomalies and large $B^+\to K^+\nu\bar\nu$ in
non-universal $U(1)'$ models}, 
arXiv:2308.13426. 
  
\bibitem{2023hiller}
R.~Bause, H.~Gisbert, and  G.~Hiller,
{\it Implications of an enhanced $B\to K\nu\bar\nu$ branching ratio}, 
arXiv:2309.00075. 

\bibitem{2023aa}
T.~Felkl, A.~Giri, R.~Mohanta, M.~A.~Schmidt, 
{\it When Energy Goes Missing: New Physics in $b\to s\nu\nu$ with Sterile Neutrinos}, 
Eur.~Phys.~J.~{\bf C 83}, 1135 (2023) [arXiv:2309.02940 [hep-ph]].

\bibitem{2023dm}  
 M.~Abdughani and Y.~Reyimuaji, 
{\it Constraining light dark matter and mediator with $B^+\to K^+\nu\bar\nu$ data}, 
arXiv:2309.03706. 

\bibitem{2023b}
  H.~K. Dreiner, J.~Y.~G\"unther, and Z.~S.~Wang,  
{\it The Decay $B\to K\nu\bar\nu$ at Belle II and a Massless Bino in R-parity-violating Supersymmetry}, 
arXiv:2309.03727.

\bibitem{2023c}
X.-G. He, X.-D. Ma, and G.~Valencia, 
{\it Revisiting models that enhance $B\to K\nu\bar\nu$ in light of the new Belle II measurement},
arXiv:2309.12741.



\bibitem{Leffa}
B.~Batell, M.~Pospelov, and A.~Ritz,
{\it Multi-lepton Signatures of a Hidden Sector in Rare B Decays},
Phys.~Rev. {\bf D83}, 054005 (2011).

\bibitem{Leffb}
  K.~Schmidt-Hoberg, F.~Staub, and M.~W.~Winkler,
  {\it Constraints on light mediators: confronting dark matter searches with B physics},
  Phys.~Lett.~{\bf B727}, 506 (2013).

\bibitem{wsb}
M.~Wirbel, B.~Stech, and M.~Bauer, {\it Exclusive Semileptonic Decays of Heavy Mesons}, 
Z.~Phys.~{\bf C29}, 637 (1985).

\bibitem{ballP}
P.~Ball and R.~Zwicky,
{\it New results on $B \to \pi, K, \eta$ decay formfactors from LCSRs},
Phys.~Rev.~{\bf D71}, 014015 (2005). 

\bibitem{ballV}
P.~Ball and R.~Zwicky,
{\it $B_{d,s} \to  \rho, \omega, K^*, \phi$ decay form-factors from LCSRs revisited}, 
Phys.~Rev.~{\bf D71}, 014029 (2005). 

\bibitem{ms2000}
D.~Melikhov and B.Stech, 
{\it Weak form-factors for heavy meson decays: an update},  
Phys.~Rev.~{\bf D62}, 014006 (2000).

\bibitem{damir2023epjc}
D.~Becirevic,  G.~Piazza, and O.~Sumensari,
{\it Revisiting $B\to K^{(*)}\nu\bar\nu$ decays in the Standard Model and beyond},
Eur.~Phys.~J.~{\bf C83}, 252 (2023).


\bibitem{hpqcd2023}
W.~G.~Parrott, C.~Bouchard, and C.~T.~H.~Davies [HPQCD collaboration], 
{\it Standard Model predictions for $B\to Kl^+l^-$, $B\to K l_1^-l_2^+$, and $B\to K \nu\bar\nu$ 
using form factors from $N_f = 2 + 1 + 1$ lattice QCD},
Phys.~Rev.~{\bf D107}, 014511 (2023); Phys.~Rev.~{\bf D107}, 119903(E) (2023).
 
\bibitem{belle2017}
J.~Grygier \textit{et al.} [Belle Collaboration]],
{\it Search for $\boldsymbol{B\to h\nu\bar{\nu}}$ decays with semileptonic tagging at Belle},
Phys.~Rev.~{\bf D96}, 091101 (2017). 

\bibitem{colangelo1997}
P.~Colangelo, F.~De Fazio, P.~Santorelli, and E.~Scrimieri, 
{\it Rare $B\to (K.K^*)\bar \nu\bar\nu$ decays at $B$-factories}, 
Phys.~Lett.~{\bf B395}, 339 (1997). 

\bibitem{mns1997a}
D.~Melikhov, N.~Nikitin, and S.~Simula,
{\it Rare decays $B\to (K, K^*) (l^+l^-,\nu\bar \nu)$ in the quark model}, 
Phys.~Lett.~{\bf B410}, 290 (1997). 

\bibitem{mns1997b} 
D.~Melikhov, N.~Nikitin, and S.~Simula,
{\it Right-handed currents in rare exclusive $B\to (K,K^*)\nu\bar\nu$ decays}, Phys.~Lett.~{\bf B428}, 171 (1997).
  
\bibitem{mns1998}
  D.~Melikhov, N.~Nikitin, and S.~Simula, 
{\it Rare exclusive semileptonic $b\to s$ transitions in the SM}, 
Phys.~Rev.~{\bf D57}, 6814 (1998). 

\end{thebibliography}
\end{document}